# Phase field simulation of dendrite growth in solid-state lithium batteries based on mechanical-thermo-electrochemical coupling[*]

HOU Pengyang[1], XIE Jiamiao[1], LI Jingyang[1, 2], ZHANG Peng[3], LI Zhaokai[4], HAO Wenqian[1, *],

TIAN Jia[1], WANG Zhe[1], LI Fuzheng[1]

1. School of Aeronautics and Astronautics, North University of China, Taiyuan 030051, China

2. Beijing Tsing Aero Armament Technology Co., Ltd., Beijing 102100, China

3. School of Mechanical and Electrical Engineering, North University of China, Taiyuan 030051,

China

4. Beijing Institute of Mechanical Equipment, Beijing 100854, China

**Abstract**：Solid-state lithium batteries possess numerous advantages, such as high energy density, excellent cycle stability, superior mechanical strength, non-flammability, enhanced safety, and extended service life. These characteristics make them highly suitable for applications in aerospace, new energy vehicles, and portable electronic devices. However, the growth of lithium dendrite at the electrode/electrolyte interface remains a critical challenge, limiting both performance and safety. The growth of lithium dendrites in the electrolyte not only reduces the Coulombic efficiency of the battery but also poses a risk of puncturing the electrolyte, leading to internal short circuits between the anode and cathode. This study is to solve the problem of lithium dendrite growth in solid-state lithium batteries by employing phase-field theory for numerical simulations. A phase-field model is developed by coupling the mechanical stress field, thermal field, and electrochemical field, to investigate the morphology and evolution of lithium dendrites under the condition of different ambient temperatures, external pressures, and their combined effects. The results indicate that higher temperature and greater external pressure significantly suppress lithium dendrite growth, leading to fewer side branches, smoother surfaces, and more uniform electrochemical deposition. Increased external pressure inhibits longitudinal dendrite growth, resulting in a compressed morphology with higher compactness, but at the cost of increased mechanical instability. Similarly, elevated ambient temperature enhances lithium-ion diffusion and reaction rate, which further suppresses dendrite growth rate and size. The combined effect of temperature and pressure exhibits a pronounced inhibitory influence on dendrite growth, with stress concentrating at the dendrite roots. This stress distribution promotes lateral growth, facilitating the formation of flatter and denser lithium deposits.

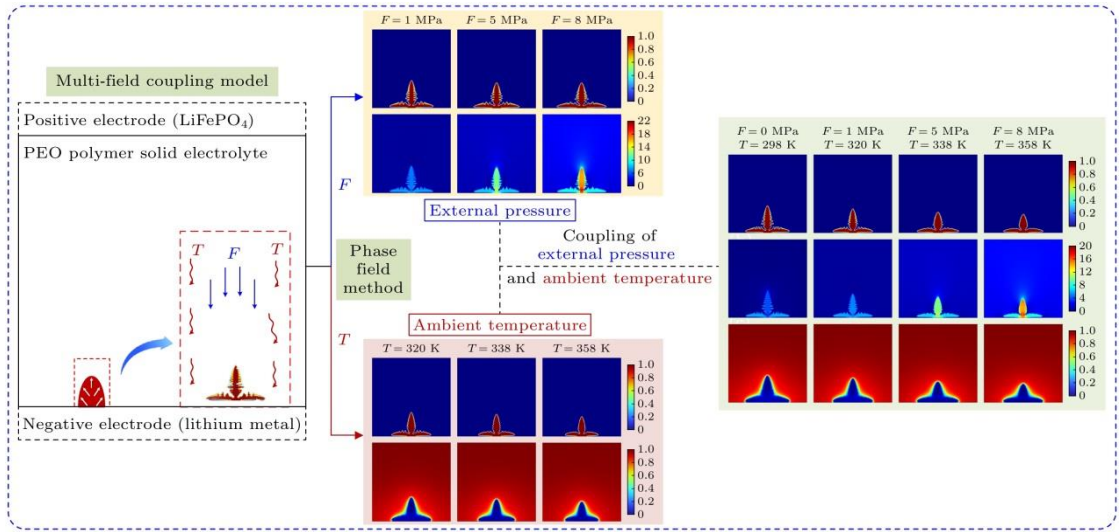

# 1. Introduction

Lithium metal has become a potential anode material because of its high theoretical capacity, low density and low redox potential, has been applied and gradually popularized in many fields and industries. In recent years, the use of lithium-ion batteries worldwide has been increasing year by year. Lithium-ion batteries play a dominant role in new energy vehicles, aerospace and other fields, profoundly affect the technological innovation and development trend of related industries, and are of great significance to promote global energy transformation and the progress of high-end manufacturing industry. Compared with conventional liquid-electrolyte lithium batteries, the solid-state lithium batteries have significantly improved[1,2] in terms of safety and electrochemical performance, and has better thermal stability, mechanical stability and higher battery energy density due to the use of solid-state electrolyte that feature non-leakage, non-volatility, non-corrosiveness, non-flammability, and high stability[3]. Therefore, as an important energy storage device, solid-state lithium batteries are of great significance in developing the national economy, reducing carbon emissions, reducing dependence on fossil energy, and addressing environmental pollution, and are regarded as one of the most promising next-generation energy storage technologies [4,5]. However, the problem of lithium dendrite growth has been a key factor restricting their performance improvement and safe application. Lithium dendrite growth not only reduces the coulombic efficiency of the battery, but also may pierce the separator and cause internal short circuits of the battery, thus causing serious safety accidents. In addition, dendrites not only destroy the solid electrolyte interface (SEI) and accelerate the side reactions at the electrode-electrolyte interface, but also lead to the formation of dead lithium, which continuously depletes active lithium and greatly reduces the service life of batteries[6,7]. Therefore, it is necessary to investigate the effect of solid-state electrolytes on lithium dendrite growth.

Lithium dendrites can be investigated by a variety of experimental methods, such as in situ or ex situ microscopy, optical microscopy, and electron microscopy. However, due to the limitations of experimental equipment, the study of lithium dendrites can only be carried out at the macro level, and it is difficult to go deep into the micro scale to capture its reaction process. In addition, the experimental method also has the problems of high time cost and highly sensitive to environmental conditions, which has led some researchers to investigate lithium dendrites using numerical simulation methods. The phase-field method is a numerical simulation method widely used in materials science, physics, chemistry and other fields. It is based on the Ginzburg-Landau theory. By introducing one or more continuously changing phase-field variables, it describes the state and phase change process of each phase in the material system with the help of differential equations based on the diffuse interface in the heterogeneous system. This method can overcome the problem of tracking the moving interface between the negative electrode and the electrolyte in the electrochemical deposition process[8]. These phase-field variables can be physical quantities such as concentration $c$, temperature $T$, external force $F$, or order parameters describing phase structure or composition[9–11]. In the aspect of simulating the growth of lithium dendrites, by solving the phase-field equations, the microscopic mechanism of lithium dendrite growth can be revealed, such as the uneven deposition of lithium ions on the electrode surface, the stress distribution inside the battery, and the growth morphology and evolution law of lithium dendrites under different internal/external conditions can also be revealed, so as to predict the growth trend of lithium dendrites, and ultimately providing theoretical guidance for battery design and optimization.

Both domestic and foreign scholars have carried out extensive studies on the dendrite growth of lithium batteries. Guyer et al. [12] first applied the phase-field method to the study of lithium dendrites, and constructed an electrochemical system model based on the one-dimensional phase-field model. Kobayashi[13] simulated the dendrite growth using the phase-field method. Both of them compared the simulation results with the experimental results, and verified the consistency between the experimental results and the simulation results. Liang et al.[14] established a one-dimensional nonlinear phase-field model to describe the evolution of the electrode/electrolyte interface in a highly non-equilibrium electrochemical process, and verified that the model could effectively predict the interface motion and microstructure evolution in a high overpotential electrochemical process, and is consistent with the classical Butler-

Volmer kinetics. Chen et al.[15] developed a nonlinear phase-field model for the lithium dendrites in liquid electrolytes . Through the analysis of three dendrite morphologies, it was found that the large applied voltage or flat protrusions at the interface would lead to the formation of dendrite side branches, and even lead to unstable tip-splitting. Based on this model, Shen et al.[16] proposed a phase-field model for lithium dendrites in solid electrolytes, and studied the effects of different external pressures on their growth behavior. The results showed that the external pressure inhibited the progress of electroplating reactions, which is not conducive to the rate performance of the battery, and will make the morphology of lithium dendritic smooth and increase its mechanical instability. Yan et al.[17] coupled the nonlinear phase-field lithium dendrite model[15] with the heat transfer model to further investigate the thermal effect of lithium dendrites, and found that an increase in ambient temperature reduced the normalized dendrite length, and the temperature gradient would prevent the formation of side branches. The uneven distribution of internal heat made the dendrite morphology change from tree shape to rhombic shape On this basis, Hong and Viswanathan[18] further proposed a fully thermally coupled phase-field model for electrochemical deposition, and found that a smaller electrochemical reaction barrier is beneficial to the inhibition of dendrite growth. Yurkiv et al.[19] developed an electrochemical–mechanical coupled phase-field model to investigate the effect of surface energy on dendrite morphology, revealing that either a low charging rate or a high interfacial $Li^+$ diffusion capacity favors the maintenance of interfacial stability. Qi et al.[20] and Arguello et al.[21] further simulated the lithium dendrite on the three-dimensional scale, and found that the increased anisotropy strength would increase the maximum height of the dendrites, while a higher temperature would help the uniform deposition of the lithium dendrites and effectively inhibit their growth, and the formation of the dendrite was related to the competition between the lithium cation diffusion and electric migration forces, such as a larger electric field near the tip of the dendritic crystal would lead to a higher lithium-ion concentration, thus triggering the tip growth of dendritic lithium. Jiang et al.[22] established a mechanical-electro-chemical coupling phase-field model of dendrite growth and crack propagation in solid electrolyte, and found that increasing the mechanical strength of solid electrolyte could slow down the dendrite growth rate and increase the stress at the dendrite tip, but did not study the influence of thermal effect on dendrite. Liang et al.[23] simulated the growth process of dead lithium and found that a larger number of charge-discharge cycles could slow down the growth of lithium dendrites during charge and reduce the area of lithium deposition, but also increase the area of dead lithium during discharge. However, higher ambient temperature can inhibit the growth of lithium dendrite and reduce the formation of dead lithium.

At present, the numerical simulation of lithium dendrite is mainly based on the single physical field or on the coupling of two physical fields [8], whereas the interactions among different influencing factors, particularly from the perspective of mechanical–thermal–electrochemical coupling, have rarely been investigated. Therefore, it is necessary to establish a lithium dendrite growth model coupling mechanical field, thermal field and electrochemical field to explore the dynamic growth process of lithium dendrites. Based on this, this study systematically investigates the effects of external pressure and temperature on the growth behavior of lithium dendrites by constructing a phase-field model of lithium dendrite in solid electrolyte and using the COMSOL Multiphysics finite element software. The regulatory mechanism underlying the morphological evolution and growth of lithium dendrites is revealed by parametric discussion for different combinations of external pressure and temperature. The coupling effect of external physical fields on lithium metal deposition kinetics and dendrite growth characteristics is discussed, and the governing roles of key influencing factors are clarified. The research focuses on the growth mode of lithium dendrite under multi-field coupling, including the regulatory effect of external pressure on interface stability and stress distribution, and the influence of different ambient temperatures on lithium-ion diffusion rates with the addition of barriers. The results not only reveal the growth law and difference of lithium dendrite under different parameters, but also provide theoretical guidance and data support for further optimization of lithium metal battery design. Through the quantitative analysis of lithium dendrite growth mechanism, the purpose is to explore effective dendrite suppression strategies to improve the safety and electrochemical performance of lithium metal batteries, and to provide a reliable technical basis for the development of next-generation high-energy-density energy storage devices.

## 2. Mechanical-thermal-electrochemical coupling model

### 2.1 Phase field evolution equation

Based on the phase-field method proposed by Chen et al.[15], a mechanical-electrochemical coupling model was established to simulate the growth of lithium dendrite in solid electrolyte by incorporating a mechanical field. During the charging process, lithium ions diffuse to the negative electrode through the solid electrolyte and combine with electrons at the lithium metal interface to form lithium. The phase-field model is constructed on the basis of the Gibbs free energy of the system. The total Gibbs free energy for lithium dendrite growth in solid electrolytes can be expressed as:

$$F = \int_V \left[ f_{ch}(\xi, c_i) + f_{grad}(\xi) + f_{elec}(\xi, c_i, \varphi) + f_{els}(\xi, u) \right] dV, \tag{1}$$

where $f_{ch}(\xi, c_i)$ represents the chemical free energy density, $f_{grad}(\xi)$ is the gradient free energy density, $f_{elec}(\xi, c_i, \varphi)$ is the electrostatic energy density, $f_{els}(\xi, u)$ is the elastic free energy density and the four variables $\xi$, $c_i$, $\varphi$, $u$ represent the phase field order parameter of lithium dendrite growth interface, mole fraction of chemical species $i$ ($i$=Li, Li$^+$, A$^-$), system potential and displacement, respectively. In the phase-field model, the order parameter $\xi$ is used as a non-conserved variable to track the evolution of lithium dendrite growth interface, which ranges from 0 to 1. When $\xi= 1$, it represents the lithium electrode phase, and when $\xi= 0$, it represents the electrolyte phase.

The chemical free energy density $f_{ch}(\xi, c_i)$ is expressed as

$$f_{ch}(\xi, c_i) = g(\xi) + RT \left( c_{Li^+} \ln\left(\frac{c_{Li^+}}{c_0}\right) + c_{anion} \ln\left(\frac{c_{anion}}{c_0}\right) \right) + \sum c_i \mu_i^\Theta, \tag{2}$$

where $R$ is the molar gas constant, $T$ is the temperature, $\mu_i^\Theta$ is the reference chemical potential of the species $i$, and $g(\xi) = W\xi^2(1-\xi)^2$ represents an arbitrary doble well function. $c_{Li^+}$ represents the concentration of Li$^+$ in the electrolyte, $c_{anion}$ represents the concentration of anions in the electrolyte, $c_0$ is the initial concentration of the electrolyte, $c_{Li^+}/c_0$ and $c_{anion}/c_0$ represent the normalized concentrations of Li$^+$ and anions, respectively.

The gradient free energy density $f_{grad}(\xi)$ is expressed as

$$f_{grad}(\xi) = \frac{1}{2}k \cdot (\nabla \xi)^2 , \qquad (3)$$

where $k = k_0[1+\delta\cos(\omega\theta)]$, $k_0$ is the gradient energy coefficient, $\delta$ is the anisotropic strength, $\omega$ represents the mode of anisotropy, and $\theta$ is the angle between the normal vector of the interface and the reference axis, representing local inhomogeneity in the system caused by metal anode defects and incomplete electrode-electrolyte contact. The electrostatic energy density $f_{elec}(\xi, c_i, \varphi)$ is expressed as

$$f_{elec}(\xi, c_i, \varphi) = \sum nFc_i\varphi , \qquad (4)$$

where $n$ represents valance of species $i$, $n = 1$ for lithium ions, and $F$ is Faraday's constant.

The elastic free energy density $f_{els}(\xi, u)$ is expressed as

$$f_{els}(\xi, u) = \frac{1}{2}C_{ijkl}\varepsilon_{ij}^E\varepsilon_{kl}^E , \qquad (5)$$

where $\varepsilon_{ij}^E = \varepsilon_{ij}^T \lambda_i h(\xi)\delta_{ij}$ represents the elastic strain tensor, $\varepsilon_{ij}^T$ is the total strain, $\lambda_i$ is the Vegard strain coefficient, and $h(\xi) = \xi^3(6\xi^2 - 15\xi + 10)$ is the interpolation function. $C_{ijkl}$ denotes the local elastic stiffness tensor, which for a single material can be written as

$$C_{ijkl} = \frac{E}{2(1+v)}(\delta_{il}\delta_{jk} + \delta_{ik}\delta_{jl}) + \frac{Ev}{(1+v)(1-2v)}\delta_{ij}\delta_{kl} , \qquad (6)$$

For the lithium metal/solid-state electrolyte composite, the local elastic stiffness tensor can also be written as

$$C_{ijkl} = \frac{E^{eff}}{2(1+v^{eff})}(\delta_{il}\delta_{jk} + \delta_{ik}\delta_{jl}) + \frac{E^{eff}v^{eff}}{(1+v^{eff})(1-2v^{eff})}\delta_{ij}\delta_{kl} , \qquad (7)$$

The elastic modulus and Poisson's ratio of the lithium metal/solid-state electrolyte composite are computed, yielding the effective elastic modulus

$E^{\text{eff}} = E^{\text{e}} h(\xi) + E^{\text{s}}\left[1 - h(\xi)\right]$ and the effective Poisson's ratio $v^{\text{eff}} = v^{\text{e}} h(\xi) + v^{\text{s}}\left(1 - h(\xi)\right)$. Where $E^{\text{e}}$, $v^{\text{e}}$ represent the elastic modulus and Poisson's ratio of lithium metal, respectively. $E^{\text{s}}$, $v^{\text{s}}$ represent the elastic modulus and Poisson's ratio of electrolyte, respectively, and $\delta_{il}\delta_{jk}$ is the Kronecker delta function.

When the phase field is used to describe the deposition process of lithium dendrite[24,25], the evolution of electrode/electrolyte interface can be described using the non-conserved variable equation, which can be expressed as

$$\frac{\partial \xi}{\partial t} = -L_i \nabla \frac{\delta F_1}{\delta \xi} - \Gamma \quad, \tag{8}$$

where $L_i$ is the kinetic coefficient and $F_1$ is the Gibbs free energy. $\Gamma$ is the electrochemical deposition rate due to the electrochemical reaction, which can be obtained by the Butler-Volmer equation

$$\Gamma = i^0 h'(\xi) \left\{ \exp\left[\frac{(1-a_1)n_1 F \eta_a}{RT}\right] - \tilde{c}_{\text{Li}^+} \exp\left[\frac{-a_1 n_1 F \eta_a}{RT}\right] \right\} \quad, \tag{9}$$

where $i^0$ is a constant related to the reaction, $n_1$ represents the number of electrons participating in the reaction, $a_1$ is the symmetry factor, and $\eta_a$ is the activation overpotential.

The evolution equation of the order parameter driven by the electrochemical reaction is obtained from Eq. (8) and Eq. (9), representing the governing equation for lithium dendrite growth:

$$\frac{\partial \xi}{\partial t} = -L_\sigma \left(g'(\xi) + f'_{\text{grad}}(\xi) + f'_{\text{els}}(\xi)\right) - L_\eta h'(\xi) \times \left\{ \exp\left[\frac{(1-a_1)n_1 F \eta_a}{RT}\right] - \tilde{c}_{\text{Li}^+} \exp\left[\frac{-a_1 n_1 F \eta_a}{RT}\right] \right\}$$

$$, \tag{10}$$

where $L_\sigma$ represents the interfacial mobility and $L_\eta$ represents the reaction constant.

The diffusion of Li$^+$ in the electrolyte is expressed by the Nernst-Planck equation[15]:

$$\frac{\partial \tilde{c}_{\text{Li}^+}}{\partial t} = \nabla \cdot \left[ D^{\text{eff}} \nabla c_{\text{Li}^+} + D^{\text{eff}} \frac{zF\tilde{c}_{\text{Li}^+}}{RT} \nabla \phi \right] - K \frac{\partial \xi}{\partial t} \quad, \tag{11}$$

where $D^{\text{eff}} = D^e h(\xi) + D^s [1-h(\xi)]$ is the effective diffusivity, $D^e, D^s$ are the diffusion coefficient of the electrode and electrolyte, respectively, and $K = 0.025$ is the ratio constant.

For the electric potential distribution, assuming charge neutrality in the system, the current density can be described by Poisson's equation, consisting of a source term $I_R$, representing the charge leaving or entering due to the electrochemical reaction:

$$\nabla \cdot [\sigma^{\text{eff}} \nabla \varphi(r,t)] = I_R = nF\rho_s \frac{\partial \xi}{\partial t} , \tag{12}$$

where $\sigma^{\text{eff}} = \sigma^e h(\xi) + \sigma^s [1-h(\xi)]$ is the effective conductivity, $\sigma^e, \sigma^s$ are the conductivities of the electrode and electrolyte, respectively, $\rho_s$ is the point concentration of lithium metal, and $I_R$ only occurs when the system deviates from equilibrium state. The lithium battery system is in a state of mechanical equilibrium, which can be expressed as

$$\nabla \cdot (C_{ijkl} \varepsilon_{kl}^E) = 0 , \tag{13}$$

$$\varepsilon_{kl}^E = \varepsilon_{kl}^T - \lambda_i h(\xi) \delta_{ij} , \tag{14}$$

$$\varepsilon^T = \frac{1}{2}[(\nabla u)^T + \nabla u] , \tag{15}$$

where $\varepsilon_{kl}^T$ represents the total strain, $\lambda_i h(\xi) \delta_{ij}$ represents the local eigenstrain induced by chemical expansion during lithium-ion insertion, and the elastic strain tensor $\varepsilon_{kl}^T$ is the difference between the total strain and the eigenstrain. Hydrostatic pressure, principal pressure and von Mises stress are used to visualize the stress evolution[16,26].

2.2 Heat transfer model

Temperature influences the growth of lithium dendrites on the electrode through its effect on the diffusivity of Li$^+$ in the electrolyte. To investigate the effect of internal heat generation on lithium dendrite growth, it is necessary to examine the spatiotemporal variation of the temperature [17]. The temperature change of the cell is usually caused by internal heat generation and the heat transfer from the surrounding environment. Therefore, the governing equation for the temperature field is :

$$C_p \rho \frac{\partial T}{\partial t} = R_a \left( \frac{\partial^2 T}{\partial x^2} + \frac{\partial^2 T}{\partial y^2} \right) + Q \ , \tag{16}$$

$$R_a = R_e h(\xi) + R_s \left[ 1 - h(\xi) \right] \ , \tag{17}$$

where $C_p$ is the specific heat capacity per unit volume, and $\rho$ is the mass density, both of which are related to the characteristics of the electrode and electrolyte. $R_a$ is the effective thermal conductivity, $R_e$ and $R_s$ are the thermal conductivities of the electrode and electrolyte, respectively. $Q$ denotes the internal heat generation rate, which includes both reversible and irreversible heat. The effect of reversible heat is negligible compared to irreversible heat[27]. Accordingly, the irreversible heat dominates the internal heat generation rate, which is expressed as

$$Q = q_{ohmic} + q_{over} \ , \tag{18}$$

$$q_{ohmic} = \sigma^e \nabla D^e \cdot \nabla D^e + \sigma^s \nabla D^s \cdot \nabla D^s \ , \tag{19}$$

$$q_{over} = a_s j \left( \varphi_e - \varphi_s - U_j \right) \ , \tag{20}$$

$$j = zF v_e \ , \tag{21}$$

where $q_{ohmic}$ is the Joule heat due to ionic resistance, and $q_{over}$ is the Joule heat due to overpotential. $D^e$ represents the diffusion coefficient of Li$^+$ in the electrode, and $D^s$ represents the diffusion coefficient of Li$^+$ in the electrolyte. The temperature dependence of these diffusion coefficients establishes the link between temperature and lithium dendrite growth, thereby achieving the coupling between the temperature field and the phase-field model. Doyle et al.[28] estimated the specific surface area per unit volume of the electrode $a_s$ by assuming that the electrode particles are spherical. $\varphi_e$ and $\varphi_s$ are the electrostatic potentials of the electrode and electrolyte, respectively. $U_j$ is the open-circuit potential, which can be regarded as a constant, and $j$ is the current density. The system is initially subjected to the condition $T = T_0$, where $T_0$ is the ambient temperature. At each time step, the radiation boundary condition is updated and applied to the system's outer boundary, and the temperature governing equation can be expressed as:

$$R_a \frac{\partial T}{\partial x} n_x + R_a \frac{\partial T}{\partial y} n_y = h(T_0 - T) + \varepsilon_R \sigma_R (T_0^4 - T^4) \ , \tag{22}$$

where $n_x$ and $n_y$ represent the two directional cosine values of the outward normal vector at the boundaries, and $h = 10$ W/(m$^2$·K) is the convective heat transfer coefficient. Define the initial temperature $T_0 = 298$ K. $\varepsilon_R = 0.49$, $\sigma_R = 5.67 \times 10^{-8}$ W/m$^2$·K$^4$ are the emission rate and Stefan-Boltzmann constant, respectively.

2.3 Coupling of heat transfer model and phase field model

Since the coupling between the heat transfer model and the phase-field model relies on determining the dependence of the Li$^+$ diffusion coefficient on temperature, this coupling is established through the functional relationship between the Li$^+$ diffusion coefficient and the temperature field:

$$D = p^e \exp\left[-bC_+ + \frac{E_D}{R}\left(\frac{1}{T_0} - \frac{1}{T}\right)\right] \ , \tag{23}$$

where $p^e = 2.582 \times 10^{-9}$ m$^2$/s and $E_D$ represent the temperature-independent pre-exponential factor and activation energy[29], respectively, and $b = 2.735 \times 10^3$ mol/m$^3$ is the fitting factor of the experimental data.

The temperature distribution inside the battery is primarily governed by the external environment as well as the internal heat generation and dissipation conditions. Temperature significantly affects the nucleation and growth of lithium dendrites through its impact on ion transport, reaction kinetics, and diffusion processes. To couple the thermal effect, the Arrhenius equation can be used to quantify the temperature sensitive parameter $X$ ($X = D, L_\eta$). Hong and Viswanathan [18] provided a more detailed heat equation based on reaction kinetics and supplemented the heat transfer equation. Combined with the research of Zhang et al.[30], the following can be obtained:

$$X_T = X_{T_{ref}} \exp\left[\frac{E_{a,X}}{R}\left(\frac{1}{T_{ref}} - \frac{1}{T}\right)\right] \ , \tag{24}$$

where $X_T$ and $X_{T_{ref}}$ represent the values of the temperature-sensitive parameter when the temperature is at $T$ and the reference temperature is at $T_{ref}$, respectively. $E_{a,X}$ is the activation energy corresponding to the parameter $X$. In order to analyze the effect of temperature on lithium dendrite more accurately, the lithium-ion diffusion barrier $E_{a,D_{Li}^+}$ and the electrochemical reaction barrier $E_{a,L_\eta}$ are introduced.

## 3. Finite element model

A nonlinear phase-field model is developed to simulate the growth of micron-sized lithium dendrites using COMSOL Multiphysics software. Since dendrite growth in solid-state lithium batteries is a localized process, which is negligible compared to the overall cell geometry, a two-dimensional square electrolyte model of 8 μm×8 μm is constructed to capture the dendrite growth in the electrolyte and the associated physicochemical processes. As shown in Fig. 1, the bottom boundary corresponds to the Li metal anode, the top boundary to the LiFePO$_4$ cathode, and the intermediate region to the PEO-based solid-state electrolyte. The initial Li$^+$ concentration at the top boundary of the electrolyte is set to $c_0$. When a potential of 0.1 V is applied to at the top boundary while no potential is applied at the bottom boundary, lithium dendrite formation can be induced by the space-charge effect. Lithium ions migrate from the cathode to the anode under the potential difference, leading to local space-charge accumulation at the Li metal anode surface. This accumulation generates a strong electric field, which in turn drives Li$^+$ nucleation and dendritic growth. The initial nucleus is assumed to be semi-elliptical, with the short axis $a$ and half of the long axis $b$. The initial values of phase field, lithium-ion concentration and electric potential are set as step functions. The initial value of phase field is set to $\text{step1}(x^2/0.2+y^2/1.2)$; the initial value of lithium ion concentration is set to $[1-\text{step1}(x^2/0.2+y^2/1.2)] \cdot c_0$; the initial value of electric potential is set to $[1-\text{step1}(x^2/0.2+y^2/1.2)] \cdot 0.1$. In addition, the symmetry and regularity of the mesh are more efficient in numerical calculation, which is helpful for the observation and analysis of the results. In the partial differential equation module, a custom free quadrilateral mesh is employed, with a maximum cell size of 0.05 μm, a minimum cell size of 0.005 μm, a maximum cell growth rate of 1.3, a curvature factor of 0.3, and a narrow area resolution factor of 1. An implicit backward differentiation formula is employed to ensure numerical stability and accuracy. The calculation time is set to 80 s and the time step is 1 s. The phase field parameters used in the finite element model are listed in Table 1.

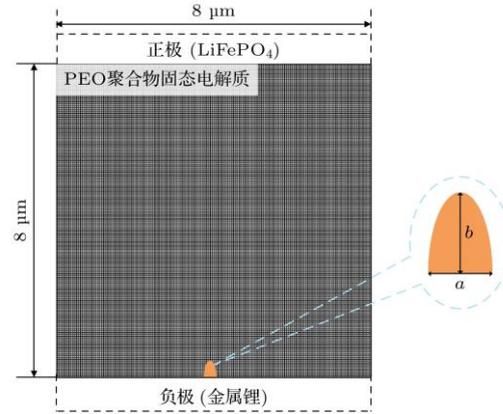

**Fig. 1.** Schematic diagram of the finite element model of two-dimensional lithium dendrite growth.

**Table 1.** Phase field simulation parameters[16,17,30,31]

| Parameter name | Parameter symbol | Numerical value | Parameter name | Parameter symbol | Numerical value |
|---|---|---|---|---|---|
| Interface mobility | $L_\sigma/(m^3 \cdot J^{-1} \cdot s^{-1})$ | $1 \times 10^{-6}$ | Reaction constant | $L_\eta/s^{-1}$ | 0.5 |
| Anisotropic strength | $\delta$ | 0.05 | Anisotropy modulus | $\omega$ | 4 |
| Electrode diffusion coefficient | $D^e/(m^2 \cdot s^{-1})$ | $1.7 \times 10^{-15}$ | Electrolyte diffusion coefficient | $D^s/(m^2 \cdot s^{-1})$ | $2 \times 10^{-15}$ |
| Electrode conductivity | $\sigma^e/(S \cdot m^{-1})$ | $1 \times 10^7$ | Electrolyte conductivity | $\sigma^s/(S \cdot m^{-1})$ | 0.1 |
| Charge transfer coefficient | $\alpha$ | 0.5 | Barrier height | $W/(J \cdot m^{-3})$ | $10^5$ |
| Initial concentration of electrolyte | $c_0/(mol \cdot m^{-3})$ | 1000 | Initial concentration of electrode | $c_s/(mol \cdot m^{-3})$ | $7.69 \times 10^4$ |
| Gradient | $k/(J \cdot m^{-1})$ | $1 \times 10^{-10}$ | Young's | $E^e/GPa$ | 7.8 |
| energy coefficient | | | modulus of electrode | | |
| Young's modulus of electrolyte | $E^s$/GPa | 1 | Poisson's ratio of electrode | $v^e$ | 0.42 |
| Poisson's ratio of electrode | $v^s$ | 0.3 | Specific heat capacity in electrode | $C_{pe}$/(J·kg$^{-1}$·K$^{-1}$) | 1200 |
| Vegard strain coefficients | $\lambda_i$ | -0.866×10$^{-3}$ | Specific heat capacity in electrolyte | $C_{ps}$/(J·kg$^{-1}$·K$^{-1}$) | 133 |
| | | -0.773×10$^{-3}$ | Thermal conductivity in electrode | $R_e$(W·m$^{-1}$·K$^{-1}$) | 1.04 |
| | | -0.529×10$^{-3}$ | Thermal conductivity in electrolyte | $R_s$/(W·m$^{-1}$·K$^{-1}$) | 0.45 |
| Li-ion diffusion barrier | $E_{a,D_{Li}^+}$/eV | 0.34 | Electrochemical reaction barrier | $E_{a,L_\eta}$/eV | 0.3 |

## 4. Results and analysis

4.1 Verification of finite element results

In order to verify the validity of the finite element model, a finite element model identical to that established by Yang et al.[31] is established, as shown in Fig. 2(a), and the results are compared with their simulations. A two-dimensional square domain of 8 μm × 8 μm was constructed to investigate the initial nucleation of lithium dendrite. The nucleation point is set in the middle of the lower boundary of the model, and the initial nucleus is set to be elliptical with $b^2/a^2 = 1.5$. During the growth process of lithium dendrite, when $t = 40$ s, the lithium-ion concentration obtained in this study (Fig. 2(b)) was generally consistent with that reported by Yang et al.[31] (Fig. 2(c)). Therefore, the present finite element model is capable of simulating the growth of lithium dendrite.

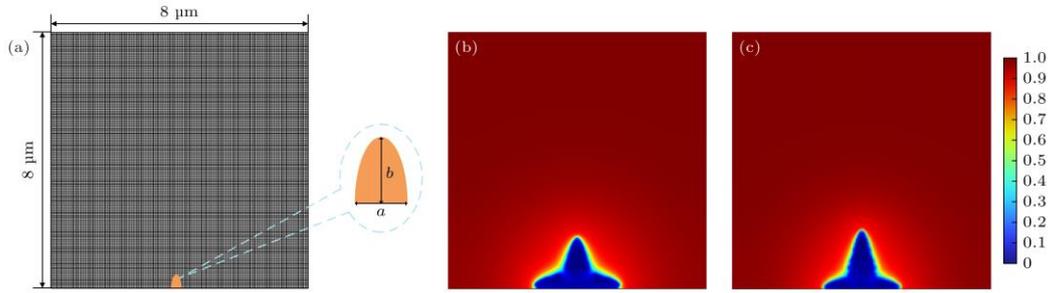

**Fig. 2.** Verification of finite element results: (a) Setting of finite element mesh and initial nucleation point; (b) lithium-ion concentration (mol/m$^3$) of current model when $t = 40$ s; (c) lithium-ion concentration (mol/m$^3$) morphology obtained by Yang et al. [31] when $t = 40$ s.

4.2 Morphology of lithium dendrite growth

When a potential of 0.1 V is applied to at the top boundary while no potential is applied at the bottom boundary, lithium dendrite formation can be induced. At room temperature ($T = 298$ K) and without external pressure ($F = 0$ MPa), the growth morphology of lithium dendrite, lithium-ion concentration, electrical potential and von Mises stress distribution are shown in the Fig. 3. Driven by the potential difference, the lithium dendrite begins to nucleate on the surface of lithium metal anode at $t = 0$ s. The accumulation of space charge at the dendrite tip forms a strong localized electric field, which promotes dendrite growth inside the solid electrolyte. In turn, the growing lithium dendrites exert pressure on the solid electrolyte, resulting in the generation of internal stress. The elastic strain energy in the solid electrolyte further drives the interfacial reaction, resulting in the continuous evolution of lithium dendrite growth. The present results are similar to those of Chen et al.[15] and Yang and Wang[31]. As shown in Fig. 3(a), after 16s of lithium deposition, pronounced dendrite growth appears at the anode interface, grows along the electrolyte while maintaining a symmetric morphology. And the growth rate and morphology of the dendrite are similar on both sides of the symmetry axis. After 80 s of growth, distinct side branches gradually emerge on the lithium dendrite, which are smaller structures branched from the trunk of the lithium dendrite. Fig. 3(b) and (c) show the corresponding lithium-ion concentration and electric potential distribution, respectively. The formation of side branches can increase the surface area of lithium dendrite and change the current distribution. Factors such as the uneven distribution of lithium in the electrolyte and the local concentration gradient result in pronounced lithium-ion concentration and potential gradients at the lithium dendrite/electrolyte interface, leading to faster dendrite growth due to the larger driving force for dendrite growth. In addition, a potential difference exists between the lithium dendrite (0 V) and the solid electrolyte (0.1 V), which also accelerates the growth rate at the interface. Fig. 3(d) illustrates the evolution of von Mises stress within the dendrite, showing a gradual increase that is concentrated at the root and bottom branches. Due to the low elastic modulus of electrolyte, its suppression effect on lithium dendrite growth is weak, and higher stress values result in greater electrolyte deformation.

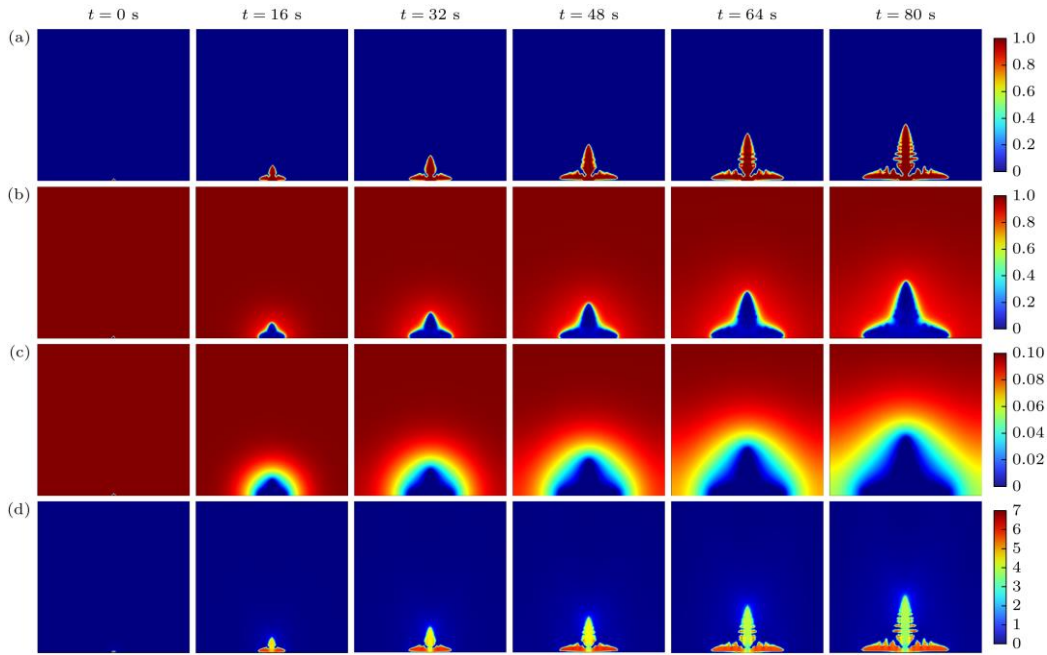

**Fig. 3.** Evolution results of lithium dendrite growth: (a) Growth morphology of lithium dendrite; (b) lithium-ion concentration (mol/m$^3$); (c) electric potential (V); (d) von Mises stress distribution (MPa).

4.3 Effect of different external pressures on lithium dendrite growth

External pressure can increase the contact area between the electrode and the electrolyte and thereby enhancing interfacial stability and suppressing the growth of lithium dendrite. The internal stress induced by the volume changes during lithium deposition and stripping, will affect lithium deposition and dissolution by affecting the diffusion of lithium ions in the electrolyte interfacial layer and the electrochemical reactions on the electrode surface[32]. In order to investigate the effect of different external pressures on lithium dendrite growth, pressures of 1 MPa, 2 MPa, 5 MPa and 8 MPa are applied on the upper boundary of the model. The Fig. 4 show the evolution of lithium dendrite growth without external pressure ($F = 0$ MPa) and with different external pressures, respectively. As shown in Fig. 4(a), with the increase of external pressure, the longitudinal growth of lithium dendrite is suppressed, while the lateral growth is promoted, exhibiting a compressive state. It is noteworthy that at $F = 2$ MPa, the dendrite height increases to $b = 2.97$ μm indicating that the suppression effect is not obvious under small external pressure, which is consistent with the findings of Shen et al.[16]. In addition, from the results of $F = 5$ MPa and $F = 8$ MPa, it is observed that the dendrite height is significantly reduced to 2.60 μm and 2.54 μm, respectively, accompanied by an increase in the main trunk width and reduction in the number of side branches. Such smooth and stocky dendrite can effectively reduce the specific surface area and improve the densification, thus providing a potential physical mechanism for stabilizing the lithium anode interface. Fig. 4(b) shows that the stress values at the trunk and root of the dendrite are larger than those in other regions. This stress concentration inhibits $Li^+$ deposition at the dendrite root, thus further regulating the growth direction of the dendritic. Comparison of stress distribution under different pressures show that stress concentration is stronger at higher pressures, which is the key factor leading to the longitudinal growth of lithium dendrite is suppressed while the lateral growth is promoted. In addition, according to the experimental study of Yin et al.[33], the external pressure also affects the $Li^+$ transport and the electrochemical reaction thereby improving battery cycling performance, which is manifested by more uniform lithium dissolution and more intensive lithium growth. When the external pressure increases gradually, the diffusion and transport of lithium ions are partially inhibited, resulting in a local decrease in $Li^+$ concentration within the dendrite. Since the growth of lithium dendrite depends on the continuous supply of lithium ions. Since dendrite growth relies on a continuous supply of $Li^+$, hindered diffusion and transport may prevent the dendrite tips and surfaces from receiving sufficient $Li^+$ to sustain further growth. In other words, the transmission channel of lithium ions is blocked due to the compression deformation of solid electrolyte.

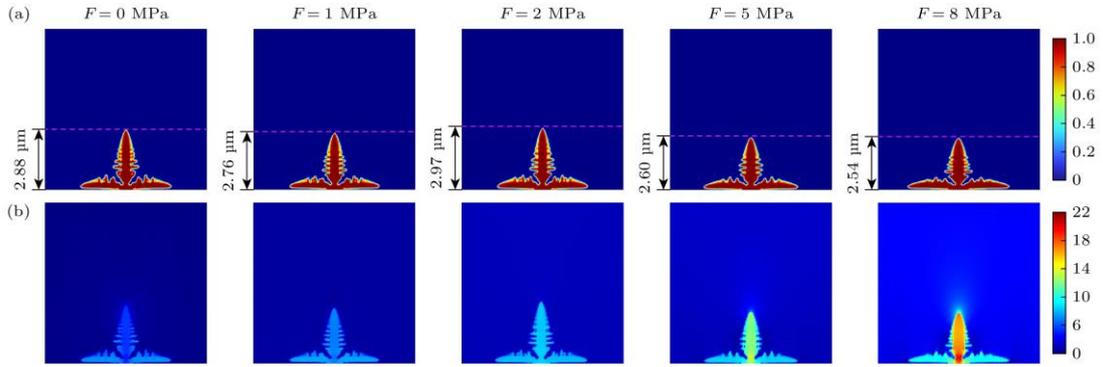

**Fig. 4.** Evolution results of lithium dendrite growth under different external pressures: (a) Growth morphology of lithium dendrite; (b) von Mises stress distribution (MPa).

Figure 5(a) shows the ratio of dendrite height to width (*b/a*) over time under different external pressures. It is observed that the morphology of lithium dendrite does not change significantly when a small pressure is applied; However, under larger pressure, the ratio decreases over time, indicating that the longitudinal growth of lithium dendrite is strongly suppressed while the lateral growth is promoted, which further confirms that larger pressure can significantly change the morphology evolution trend of lithium dendrites. Figure5(b) presents the variation of the width-to height ratio (*a/b*) under different external pressures. When the external pressure is small, the inhibition of lithium dendrite is not obvious. When the external pressure exceeds 2 MPa, the growth of lithium dendrite is suppressed. When the external pressure increases, the growth of lithium dendrite is gradually suppressed, and the growth trend in the horizontal direction is greater than that in the vertical direction. Although external pressure effectively suppresses the dendrite growth, it simultaneously the risk of material failure[34]. Once the von Mises stress exceeds the yield strength, the dendrite fails to withstand the applied stress and undergo plastic deformation or fracture, resulting in the electrolyte cannot rebound with the stripping of lithium dendrite. However, in this work, only ideal elastic strain is considered. In Fig. 4(b), when the external pressure is 8 MPa, the von Mises stress at the root of the dendrite is maximized, indicating a greater risk of fracture. The fractured lithium dendrite will lose its electrical contact with the solid electrolyte and can no longer participate in the electrochemical reactions, thus converting into dead lithium. The formation of dead lithium not only reduces the amount of active lithium but also may increase the internal resistance of the battery, further declining the coulombic efficiency. An appropriate external pressure range exerts a beneficial influence on both the electrochemical and mechanical properties of cells. The lower limit of the external pressure can ensure uniform and stable lithium deposition, whereas exceeding the upper limit renders battery performance insensitive to external pressure and may even lead to degradation [32].

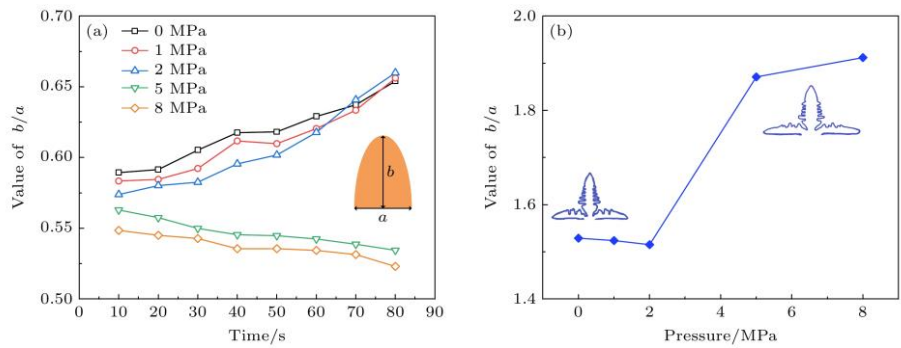

**Fig. 5.** Results of lithium dendrite growth trend under different external pressures: (a) The ratio of height to width ($b/a$) of lithium dendrite changes with time; (b) the ratio of width to height ($a/b$) of lithium dendrite changes with pressure.

4.4 Effect of ambient temperature on lithium dendrite growth

Lithium battery involves complex mechanical-thermal-electrochemical coupling changes during operation, and the effect of temperature on lithium dendrite is also complex. On the basis of the existing heat transfer model, incorporating the coupling between reaction kinetics and temperature is helpful to deeply analyze the relationship between the interaction between ion transport and electrochemical reaction rate and temperature from the microscopic point of view. Figure 6 presents the evolution of lithium dendrite morphology and lithium-ion concentration at 298 K, 320 K, 338 K and 358 K when the diffusion barrier and electrochemical barrier are combined. The Fig. 6(a) shows that at 338 K and 358 K, the dendrite height is significantly lower than that at 298 K and 320 K, and the growth rate is suppressed, but the width of the dendrite trunk increases slightly. In Fig. 6(b), the lithium-ion concentration from the electrolyte to the dendrite tip decreases slowly, and the concentration gradient at the interface between the dendrite and the electrolyte decreases, which is similar to the findings of Qiao Dongge et al.[35], Hong and Viswanathan[18]. This change can be attributed to the enhanced ion mobility at increased temperature, which reduces the local concentration gradient. The effect of high temperature on lithium dendrite morphology can be explained in many ways. First, increased temperature increases the diffusion coefficient of lithium ion and the electrochemical reaction rate, which promotes the uniform deposition of lithium-ion, lowering nucleation density, and enlarging nucleus size[36]. Second, the increase of temperature leads to the decrease of the stability of electrolyte materials, which suppresses lithium dendrite growth. When the ambient temperature reaches 358 K, this inhibitory effect becomes more obvious with increasing temperature. There are almost no side branches on the main stem of lithium dendrite, and the surface of dendrite becomes smooth. High temperature may also accelerate interfacial reactions between electrolyte and lithium dendrite, leading to higher interfacial energy and the inhibition of side branch growth of lithium dendrite[17]. Fig. 7 shows the ratio of lithium dendrite height to width ($b/a$) with time at different ambient temperatures. It can be observed that the ratio increases with time and decreases with increasing temperature. This indicates that the transverse growth of dendrites is more significant than the longitudinal growth at high temperature, which makes the morphology becomes broad and short.

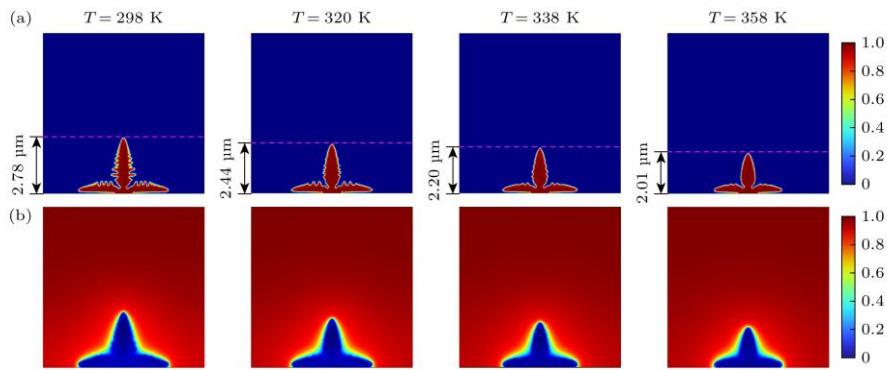

**Fig. 6.** Evolution results of lithium dendrite growth at different ambient temperatures: (a) Growth morphology of lithium dendrite; (b) lithium-ion concentration (mol/m$^3$).

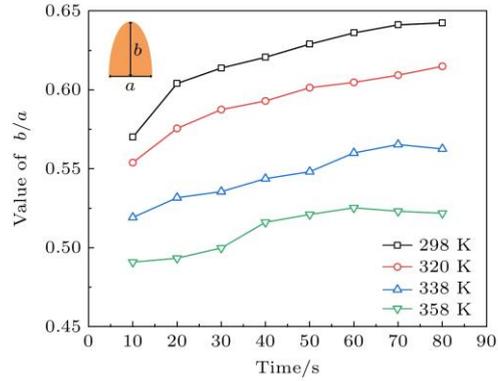

**Fig. 7.** Ratio of height to width ($b/a$) of lithium dendrite changes with time at different ambient temperatures.

4.5 Coupling effect of external pressure and temperature on lithium dendrite growth

To investigate the synergistic effect of ambient temperature and external pressure on the suppression of lithium dendrite growth, Fig. 8 presents the evolution lithium dendrite growth under the coupling effect of different ambient temperatures and external pressures. As shown in Fig. 8, when the two conditions are coupled, the suppression in solid electrolyte is more obvious than that in single condition. The combined effect of a large external pressure applied in the vertical direction ($F = 8$ MPa) and a high temperature ($T = 358$ K) significantly inhibits the vertical growth of dendrite, and the lithium dendrites grow towards the horizontal direction, thus showing an obvious cone-shaped. Fig. 8(b) shows that the von Mises stress concentration occurs at the root and trunk of the dendrite, which is conducive to the formation of flat and dense lithium deposition. This is attributed to the local stress concentration caused by the extrusion of the solid electrolyte during the dendrite growth. The applied pressure mechanically compresses the entire structure, and due to its larger volume and role as the central load-bearing region, stress is preferentially transmitted to the stem. At the dendrite tip, the sharp shape with higher curvature produces an obvious tip effect, leading to significant stress concentration and increased lithium-ion accumulation (Fig. 8(c)). Especially at high pressure ($F = 8$ MPa) and high temperature ($T = 358$ K), the von Mises stress at the root and stem is larger, and the lithium-ion accumulation at the tip becomes more concentration. Fig.9 compares dendrite growth heights under external pressure, ambient temperature and the coupling of these two conditions. Compared with the results under a single condition, the lithium dendrite growth under the two conditions shows a stronger inhibition effect on the whole, with a nonlinear trend. In the von Mises stress distribution, when comparing the stress values under coupling conditions (Fig. 8(b)) with those under only external pressure(Fig. 4(b)), the maximum von Mises stress under coupling conditions is 20 MPa, which is smaller than the 22 MPa under only external pressure. This is because the increased temperature inhibits the dendrite growth, which partially offsets the stress concentration caused by the external pressure, and the stress under the coupling condition is more dispersed to the width direction. Since temperature and electrochemical barrier reduce the overall dendrite size, the lateral growth driven by pressure is also partially offset. The dendrite height under the coupling condition is smaller than that under the only external pressure, which indicates that the coupling effect of temperature and pressure has a significant synergistic effect on the inhibition of lithium dendrite growth. The morphology and evolution of lithium dendrite growth are systematically examined under ambient temperature, external pressure and their coupled effects.

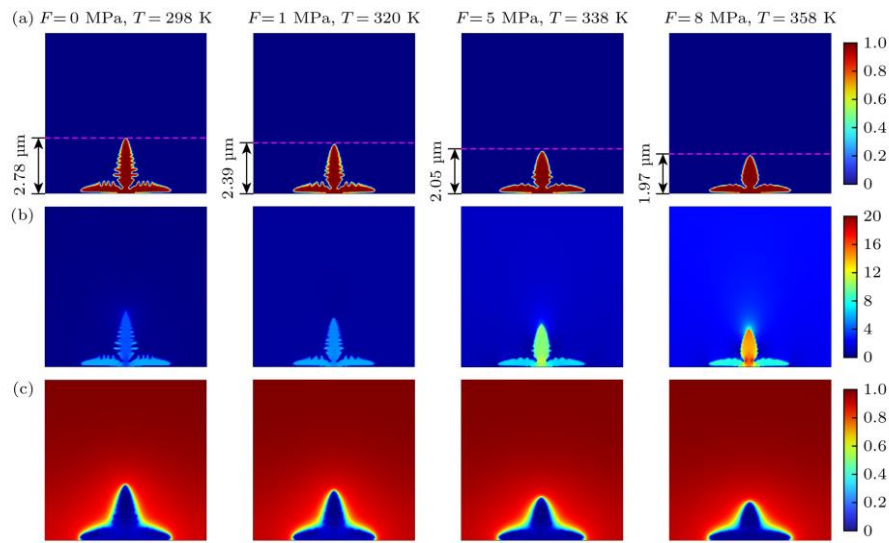

**Fig. 8.** Evolution results of lithium dendrite growth under coupling of external pressure and ambient temperature: (a) Growth morphology of lithium dendrite; (b) von Mises stress (MPa); (c) lithium-ion concentration (mol/m$^3$).

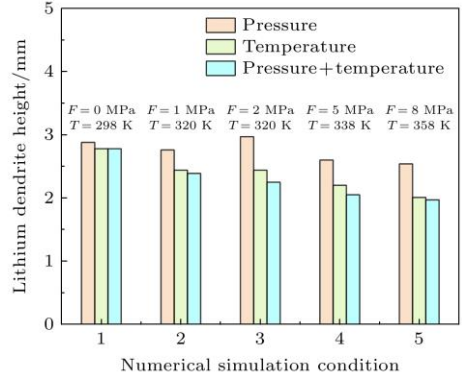

**Fig. 9.** Comparison of lithium dendrite height under external pressure, ambient temperature and coupling of these two conditions.

## 5. Conclusion

In this study, a phase-field model is developed to developed to simulate lithium dendrite growth in solid-state electrolytes. The external pressure and temperature are adjusted to investigate their effects on dendrite evolution. The coupled effects of pressure and temperature on dendrite morphology and growth evolution are systematically examined.

1) With the increase of external pressure, the growth of lithium dendrite is suppressed. Larger external pressure reduces the stress concentration area of the electrode material, which makes the deposition of lithium ions on the electrode surface more uniform and promote to reduce the formation of lithium dendrites. With the increase of external pressure, the morphology of lithium dendrite also changes. The dendrite morphology transforms from a needle-like structure to as short, thick morphology with smooth surface area and higher compactness. This morphological change is beneficial to reduce the surface area and interfacial energy of lithium dendrite, thus further inhibiting its growth. Therefore, larger external pressure facilitates uniform electrodeposition in solid-state electrolytes, inhibiting the growth of lithium dendrites. The stocky lithium dendrite structures are mechanically more stable and are less prone to fracture or detachment due to external factors, thereby improving the stability and safety of the battery. In addition, the von Mises stress at the root of the dendrite is relatively larger, resulting in stress concentration, which further affects the growth direction of the dendrite. However, attention must also be paid to the maximum stress and yield strength that the material can withstand, so as to prevent the formation of dead lithium caused by dendrite fracture at locations with high stress values.

2) When investigating the effect of ambient temperature on lithium dendrite growth, the heat transfer model is related to the diffusion coefficient and reaction constant, enabling the analysis of thermal effect on lithium dendrite growth from a more microscopic perspective. When the electrochemical reaction barrier is smaller than the diffusion barrier, increasing temperature leads to a reduction in dendrite size and a suppression of dendrite growth. Moreover, the number of side branches on the main stem of lithium dendrite decreases gradually, and the surface tends to be smooth. Under high temperature environment, the stability of electrolyte materials is significantly reduced, thereby strongly suppressing dendrite growth. Higher ambient temperature increases the diffusivity of lithium-ion, accelerating ion transport in the electrolyte and reducing interfacial concentration gradients. It also accelerates interfacial reactions and increases interfacial energy, which inhibits the growth of lithium dendrite.

3) Single factor (temperature or external pressure) can suppress dendrite growth to some extent, but when the two factors are coupled, the growth morphology of lithium dendrite may become more complex and less predictable. This suggests the necessity of a more comprehensive and systematic analysis of Multiphysics coupling effects in dendrite growth studies. Overall, larger external pressure and temperature exert a significant inhibitory effect on lithium dendrite growth. Lateral growth becomes dominant while the longitudinal growth trend is weakened, and the dendrite morphology exhibits a more obvious compressed state. This morphological change contributes to the formation of a flatter and denser lithium deposition layer, which provides favorable conditions for optimizing the performance of solid-state lithium batteries.

In this study, the stress distribution during lithium dendrite growth is simulated using the elasticity-based mechanical model, with the von Mises stress as the evaluation index. However, the definition of von Mises stress is usually used to describe the yield behavior of materials. It should be noted that the present work does not account for the plastic behavior of either lithium metal or the solid electrolyte. Instead, the stress concentration effects under different conditions are only analyzed under elastic assumption. In practical applications, lithium metal may undergo significant plastic deformation or creep in the high-stress regions, and the increase of temperature may further change its yield behavior and elastic modulus. Therefore, future work can introduce plastic mechanical behavior on the basis of the existing model, and by combining elastic and plastic constitutive relations and validating through experiments, to provide a more comprehensive description of the influence of pressure on lithium dendrite growth and stress distribution.

**References**

[1] Goodenough J B, Singh P 2015 *J. Electrochem. Soc.* **162** A2387

[2] Peters B K, Rodriguez K X, Reisberg S H, Beil S B, Hickey D P, Y Kawamata, Collins M, Starr J, Chen L, Udyavara S, Klunder K, Gorey T J, Anderson S L, Neurock M, Minteer S D, Baran P S 2019 *Science.* **363** 838

[3] Geng X B, Li D G, Xu B 2023 *Acta Phys. Sin.* **72** 220201

[4] Viswanathan V, Epstein A H, Chiang Y M, Esther T, Bradley M, Langford J, Winter M 2022 *Nature.* **601** 519

[5] Lee M J, Han J, Lee K, Lee Y J, Kim B G, Jung K N, Kim B J, Lee S W 2022 *Nature.* **601** 217

[6] Hao F, Verma A, Mukherjee P P 2018 *J. Mater. Chem. A* **6** 19664

[7] Liu Z, Qi Y, Lin Y X, Chen L, Lu P, Chen L Q 2016 *J. Electrochem. Soc.* **163** A592

[8] Zhang G, Wang Q, Sha L T, Li Y J, Wang D, Shi S Q 2020 *Acta Phys. Sin.* **69** 226401

[9] Sripad S, Viswanathan V 2017 *Electrochem. Soc.* **164** E3635

[10] Sripad S, Viswanathan V 2017 *ACS Energy Lett.* **2** 1669

[11] Guttenberg M, Sripad S, Viswanathan V 2017 *ACS Energy Lett.* **2** 2642

[12] Guyer J E, Boettinger W J, Warren J A, McFadden G B 2004 *Phys. Rev. E* **69** 021603

[13] Kobayashi R 1993 *Physica D* **63** 410

[14] Liang L Y, Qi Y, Xue F, Bhattacharya S, Harris S J, Chen L Q 2012 *Phys. Rev. E*

**86** 051609

[15] Chen L, Zhang H W, Liang L Y, Liu Z, Qi Y, Lu P, Chen J, Chen L Q 2015 *J. Power Sources* **300** 376

[16] Shen X, Zhang R, Shi P, Chen X, Zhang Q 2021 *Adv. Energy Mater.* **11** 2003416

[17] Yan H H, Bie Y H, Cui X Y, Xiong G P, Chen L 2018 *Energy Convers Manag.* **161** 193

[18] Hong Z J, Viswanathan V 2019 *ACS Energy Lett.* **4** 1012

[19] Yurkiv V, Foroozan T, Ramasubramanian A, Shahbazian-Yassar R, Mashayek F 2018 *MRS Commun.* **8** 1285

[20] Qi G Q, Liu X L, Dou R F, Wen Z, Zhou W N, Liu L 2024 *J. Energy Storage* **101** 113899

[21] Arguello M E, Labanda N A, Calo V M, Gumulya M, Utikar R, Derksen J 2022 *J. Energy Storage* **53** 104892

[22] Jiang W J, Wang Z H, Hu L Z, Wang Y, Ma Z S 2024 *J. Energy Storage* **86** 111126

[23] Liang C, Xing P F, Wu M W, Qin X P 2024 *Energy Storage Sci. Techn.* **1125** 2095

[24] Cahn J W, Allen S M 1977 *J. Phys. IV* **38** C7

[25] Allen S M, Cahn J W 1979 *Acta Metall.* **27** 1085

[26] Liang Y H, Fan L Z 2020 *Acta Phys. Sin.* **69** 226201

[27] Wu W, Xiao X, Huang X S 2012 *Electrochim. Acta* **83** 227

[28] Doyle M, Newman J, Góźdź A S, Schmutz C, Tarascon J M 1996 *J. Electrochem. Soc.* **143** 1890

[29] Stewart S G, Newman J 2008 *J. Electrochem. Soc.* **155** F13

[30] Zhang Y X, Li Y F, Shen W J, Li K, Lin Y X 2023 *ACS Appl. Energy Mater.* **6** 1933

[31] Yang H D, Wang Z J 2023 *J. Solid State Electr.* **27** 2607

[32] Cui J, Shi C, Zhao J B 2021 *CIESC J.* **72** 3511

[33] Yin X S, Tang W, Jung I D, Phua K C, Adams S., Lee S W, Zheng G W 2018 *Nano Energy* **50** 659

[34] Wang Q Y, Wang S, Zhou G, Zhang J N, Zheng J Y, Yu X Q, Li H 2018 *Acta Phys. Sin.* **67** 128501

[35] Qiao D G, Liu X L, Wen Z, Dou R F, Zhou W N 2022 *ESST* **11** 1008

[36] Yan K, Wang J Y, Zhao S Q, Zhou D, Sun B, Cui Y, Wang G X 2019 *Angew Chem. Int. Edit.* **58** 11364